\documentclass[12pt]{article}
																			 																																																												\pdfoutput=1	
\usepackage{amsmath,amssymb}
\usepackage{graphicx}
\usepackage[pdftex]{hyperref}
\usepackage{epstopdf}
\DeclareGraphicsExtensions{.eps,.bmp,.wmf,.jpeg,.pdf}
\numberwithin{equation}{section}
\def\be{\begin{equation}}
\def\ee{\end{equation}}

\def\bea{\begin{eqnarray}}
\def\eea{\end{eqnarray}}
\title{A Viable model for modified gravity}
\author{L. N. Granda\thanks{ngranda@univalle.edu.co}\\ {\small\it Departamento de Fisica, Universidad del Valle}\\{\small\it A.A. 25360, Cali, Colombia}}
\date{}
\begin{document}
\maketitle

\begin{abstract}
\noindent  We propose a model for modified gravity that meets the conditions of viability. The model has stable constant curvature solution and for an special case contains flat space time solution. The model also leads to matter stability under small perturbations of scalar curvature.  We give an example of restrictions on the parameters that  give large enough scalaron mass to avoid detectable corrections to the Newton's law. The model describes inflation and late time accelerated expansion with an effective cosmological constant for the inflationary epoch and small effective cosmological constant for the current accelerated phase. 
\\ 

\end{abstract}

\section{Introduction}
\noindent 
Among the alternatives to the explanation of dark energy (for review see \cite{copeland, sahnii, padmanabhan, sergeiod}), the large-distance modification of gravity represents an interesting alternative and has received much attention. To the large-distance modification of gravity belong the modified gravity theories that generalize the Einstein-Hilbert Lagrangian by adding some  corrections encoded in a function $f(R)$ (see \cite{sodintsov1, sotiriou, tsujikawa0, sodintsov1a} for reviews) , that may become relevant at late universe \cite{capozziello}, \cite{capozziello1}, \cite{sodintsov}, \cite{carroll}. The $f(R)$ theories have been intensively studied to explain the late time accelerated expansion and many types of modifications to the Einstein-Hilbert action have been proposed so far \cite{sodintsov1, capozziello, capozziello1, sodintsov, carroll, faraoni, dobado, anthoni, barrow1, nojiri,  elizalde, troisi, allemandi, koivisto, brevik, sodintsov2, nojiri1, nojiri2, olmo, hu}.
In addition to the possibility of explaining late-time acceleration, $f(R)$ theories may be relevant to early-time inflation at large $R$, due to its non-linear character \cite{starobinsky}. 
The first proposal to unify the inflation with late-time acceleration as well as construction of number of viable models of such acceleration is made by Nojiri and Odintsov in \cite{nojiri5, nojiri6, nojiri7, nojiri8}. 
Any realistic model of modified gravity should pass not only the local tests where the average density of matter is high compared with that of the universe, but also the observational cosmological restrictions. To pass solar system tests the model has to implement the so called chameleon mechanism \cite{hu, tsujikawa, brax} which gives a large enough mass to the scalar field (in the Einstein frame) to avoid measurable corrections to the local gravity  phenomena which is well described by the general relativity. Models that can satisfy both cosmological and local gravity constraints have been proposed in \cite{hu, astarobinsky, appleby1, sergeid1, sergeid2, eelizalde}. Exact solutions explaining the current accelerated expansion are presented in \cite{bamba1, barrow, clifton, capozz, capozz1, capozz2}.
In this paper we propose a viable model of $f(R)$ gravity that unifies early time inflation with late time accelerated expansion. It is shown that the model is free of ghosts and tachyon instabilities, satisfies local systems tests and gives an adequate description of the dark energy. The model also contains constant curvature solutions and is stable under matter perturbations.\\
This paper is organized as follows. In section II we present the model and the equations of motion in general background and in the FRW metric. In section III we derive the conditions for viability, find the constant curvature solutions and show the matter stability. In section IV we present some discussion.
\section{Field Equations}
A generalization of the Einstein-Hilbert action with the Lagrangian $R-2\Lambda$, is given by the following action
\be\label{eq1}
S=\int d^4x\sqrt{-g}\left[\frac{1}{2\kappa^2}\left(R+f(R)\right)+{\cal L}_m(\psi)\right]
\ee
where $\kappa^2=8\pi G$ and ${\cal L}_m$ is the Lagrangian density for the matter component. In what follows and when needed we will represent $F(R)=R+f(R)$. Variation with respect to the metric gives the following equation of motion
\be\label{eq2}
F'(R)R_{\mu\nu}-\frac{1}{2}g_{\mu\nu}F(R)+\left(g_{\mu\nu}\Box-\nabla_{\mu}\nabla_{\nu}\right)F'(R)=\kappa^2 T_{\mu\nu}^{(m)}
\ee
where $T^{(m)}_{\mu\nu}$ is the matter energy-momentum tensor and the prime indicates the derivative with respect to $R$. Taking the trace of eq. (\ref{eq2})  leads to 
\be\label{eq3}
RF'(R)-2F(R)+3\Box F'(R)=\kappa^2 T^{(m)}
\ee
The time and spatial components of the Eq. (\ref{eq2}) can be written in more conventional form as
\be\label{eq2a}
3H^2=-\frac{1}{2}f(R)+3\left(H^2+\dot{H}\right)f'(R)-18H\left(4H\dot{H}+\ddot{H}\right)f''(R)+\kappa^2\rho_m
\ee
and
\be\label{eq2b}
\begin{aligned}
-3H^2-2\dot{H}=&\frac{1}{2}f(R)-\left(3H^2+\dot{H}\right)f'(R)+6\left(8H^2\dot{H}+4\dot{H}^2+6H\ddot{H}+\dddot{H}\right)f''(R)\\
&+36\left(4H\dot{H}+\ddot{H}\right)^2f'''(R)+\kappa^2p_m
\end{aligned}
\ee
where all curvature dependent terms in the r.h.s. of both equations can be interpreted as effective density and pressure respectively, due to the new interaction introduced by the $f(R)$ correction.
The equation (\ref{eq3}) has an interesting interpretation if one introduces the so called "scalaron" field $F'(R)$.  By writing the  Eq. (\ref{eq3})  in the form
\be\label{eq5} 
\Box F'(R)=\frac{\partial V_{eff}}{\partial F'(R)}
\ee
where 
\be\label{eq6}
\frac{\partial V_{eff}}{\partial F'(R)}=\frac{1}{3}\left(2F(R)-RF'(R)-\kappa^2\rho_m\right)
\ee
where the trace of the energy-momentum tensor is $T^{(m)}=-(\rho_m-3p_m)=-\rho_m$ for non-relativistic matter. This effective potential has an extremum at the solution of the equation
\be\label{eq7}
 2F(R)-RF'(R)=\kappa^2\rho_m
\ee
For a given $F(R)$ and in absence of matter this is an algebraic equation on $R$, and one has a de Sitter solution associated with constant scalar curvature $R_0$ as
\be\label{eq8}
2F(R_0)-R_0 F'(R_0)=0
\ee
Evaluating the second derivative of the potential at this extrema (in absence of matter)  gives the mass of the scalaron field as
\be\label{eq9}
m^2_F=\frac{1}{3}\left(\frac{F'(R)}{F''(R)}-R\right)\Big{|}_{R=R_0}
\ee
Thus, if $m_F^2>0$, then the de sitter solution is stable.\\
There are a number of cosmological and local gravity constraints that viable $F(R)$ models should satisfy.  The first one comes from the avoidance of antigravity regime which leads to the condition $F'(R)>0$ (or $f'(R)>-1$.  This condition guarantees that the effective Newton constant, $G_{eff}=G/(1+f'(R))$ keeps the sign during all evolution. Once this condition is satisfied, the second condition related with the avoidance of tachyonic instability, demands the positivity of the mass of the scalar particle associated with $F(R)$, i.e. $F''(R)>0$. The condition $F''(R)>0$ guarantees that the evolution in the high curvature regime is stable under small perturbations \cite{appleby1}. The next constraint we consider on $F(R)$ is that in the limit $R\rightarrow \infty$ we have $F(R)/R\rightarrow 1$ (or equivalently $f(R)/R\rightarrow 0$), from which follows that $f(R)\rightarrow const.$ as $R\rightarrow \infty$ (here we assume this behavior though it is enough that $f(R)\propto R^{\sigma}$ with $\sigma<1$), which guarantees that in the large $R$ regime the $f(R)$ correction acts as a suitable cosmological constant that leads to the $\Lambda$CDM where it is well tested by the CMB. On the other hand, assuming that $F''(R)>0$ all the time, then $F'(R)$ is monotonic  increasing function, which means that $F'(R)$ tends to unity from bellow (or $f'(R)\rightarrow 0$ as $R\rightarrow \infty$) and then $F'(R)<1$ . This is summarized in the inequality $-1<f'(R)<0$. At the current epoch of low redshift ($R=R_0$) $f(R)$ should be close to the cosmological constant and satisfy the constraints from the observed accelerated expansion. In the next section we present a model that satisfy all these conditions.\\
We may interpret the model (\ref{eq1}) in the Einstein frame (which is specially useful to study the inflation) by performing a conformal transformation of the metric with the function $F'(R)$ \cite{barrow2}
\be\label{eq13}
g_{\mu\nu}\rightarrow \tilde{g}_{\mu\nu}=F'(R)g_{\mu\nu}=e^{-\sqrt{\frac{2}{3}}\kappa\phi}g_{\mu\nu},
\ee
the action takes the form
\be\label{eq14}
S=\int d^4x\sqrt{-\tilde{g}}\left[\frac{1}{2\kappa^2}\tilde{R}-\frac{1}{2}\tilde{g}^{\mu\nu}\partial_{\mu}\phi\partial_{\nu}\phi-V(\phi)+{\cal L}_m(e^{\sqrt{\frac{2}{3}}\kappa\phi}\tilde{g}_{\mu\nu},\psi)\right]
\ee
with the potential
\be\label{eq15}
V(\phi)=\frac{R(\phi)F'(R(\phi))-F(R(\phi))}{2\kappa^2 F'(R(\phi))^2}
\ee
in which the new scalar field "scalaron" couples minimally to the scalar curvature but becomes coupled to the matter sector. 
\section{A viable $f(R)$ model}
In this section we consider the following model 
\be\label{eq15a}
f(R)=-\mu^2\left(\frac{\alpha R+\beta}{\gamma R+\delta}\right)^{\eta}
\ee
where all the constants are positive, $\eta$ can be a real positive, the constants $\alpha, \gamma$ have dimensions of $length^2$ and $\beta, \delta$ are dimensionless. The energy scale $\mu^2$ has the appropriate value depending on the power $\eta$. An interesting fact of this $f(R)$ correction is that it has a "duality" symmetry under the transformation $R\rightarrow 1/R$ (and $\alpha\leftrightarrow \beta, \gamma\leftrightarrow \delta$), which could be useful in some asymptotic cases. The model has also two additional symmetries: under $(\alpha,\beta,\gamma,\delta)\rightarrow (-\alpha,-\beta,-\gamma,-\delta)$ and under $\eta\rightarrow -\eta$, $\alpha\leftrightarrow\gamma, \beta\leftrightarrow\delta$.
Applying to (\ref{eq15a}) the requirements discussed above we find
\be\label{eq15b}
\lim_{R\rightarrow\infty}f(R)=-\mu^2\left(\frac{\alpha}{\gamma}\right)^{\eta}=-2\Lambda_i
\ee
where $\Lambda_i$ is an effective cosmological constant at the early universe, which generates inflation provided that $\Lambda_i\sim 10^{46}$ ${\rm ev}^2$ (the expected typical energy of inflation). For late time cosmology, in order to generate accelerated expansion the correction $f(R)$ should be very close to the observed value of the cosmological constant
\be\label{eq15c}
f(R_0)\approx-2\Lambda
\ee
then using (\ref{eq15a}) we find the following expression for the current scalar curvature
\be
R_0=\frac{\delta(2\Lambda/\mu^2)^{1/\eta}-\beta}{\alpha-\gamma(2\Lambda/\mu^2)^{1/\eta}}
\ee
which for a given $R_0$ leads to a restriction between the parameters of the model. Note also that 
\be\label{eq15d}
\lim_{R\rightarrow 0}f(R)=-\mu^2\left(\frac{\beta}{\delta}\right)^{\eta}
\ee
which with a good approximation can be considered as the current value of the cosmological constant that is consistent with current observations of accelerated expansion, i.e. $\mu^2(\beta/\delta)^{\eta}\sim 2\Lambda\sim 10^{-66} {\rm ev}^2$. So the asymptotic behavior limits the relations between the parameters of the model.\\ 
Note that this model covers some previous models of dark energy in the limit where some of the parameters are canceled. Thus the power law models are obtained in the cases $\alpha=\delta=0$ giving $f(R)\propto 1/R^{\eta}$ and $\beta=\gamma=0$ leading to $f(R)\propto R^{\eta}$. Another interesting case is obtained by setting $\beta=0$ which leads to the model
\be\label{eq15e}
f(R)=-\mu^2\frac{(\alpha R)^{\eta}}{(\gamma R+\delta)^{\eta}},
\ee
that fulfills the condition $f(0)=0$ which means that the model has flat Minkowski space time solution corresponding to vanishing cosmological constant. For $\eta=1$ the model (\ref{eq15e}) coincides with the case $n=1$ of the model \cite{hu}. In fact if we redefine the constants $\alpha=\mu^2\tilde{\alpha}, \gamma=\mu^2\tilde{\gamma}$ in (\ref{eq15e}) and consider as in \cite{hu} relatively large curvature $R/\mu^2\sim 40$, then we can make the approximation
\be\label{eq15f}
f(R)=-\mu^2\left(\frac{\tilde{\alpha}}{\tilde{\gamma}}\right)^{\eta}\left(1+\frac{\delta}{\tilde{\gamma}}\left(\frac{\mu^2}{R}\right)\right)^{-\eta}\approx -\mu^2\left(\frac{\tilde{\alpha}}{\tilde{\gamma}}\right)^{\eta}\left(1-\frac{\eta\delta}{\tilde{\gamma}}\left(\frac{\mu^2}{R}\right)\right)
\ee
which gives a constant $-\mu^2\left(\frac{\tilde{\alpha}}{\tilde{\gamma}}\right)^{\eta}$ that responds for the current acceleration of the universe. This approximation is valid for curvatures larger than the current curvature $R_0$ (assuming $R_0/\mu^2\sim 40$) and describes the past expansion history. Neglecting $f'(R)$ at the current epoch, we have from (\ref{eq7}) (and also considering that $\Box F'(R)\approx 0$ i.e. the scalaron field is near the minimum of the potential)
\be\label{eq15g}
R\sim \kappa^2\rho_m-2f(R)\sim \kappa^2\rho_m+2\mu^2\left(\frac{\tilde{\alpha}}{\tilde{\gamma}}\right)^{\eta}
\ee
Note that as the coefficient in the second term in (\ref{eq15f}) is smaller the model becomes closer to the $\Lambda$CDM.\\
From now on we will concentrate in the case $\eta=1$ in the model (\ref{eq15a}) since this is the simplest case and allows to solve analytically some equations. For $\eta=1$  the model takes the form
\be\label{eq16}
f(R)=-M^2\frac{\alpha R+\beta}{\gamma R+\delta}
\ee
where $M^2$ is an appropriate mass scale for $\eta=1$ and all the constants are positive. This model can be considered as a modification of the Hu-Sawicki model \cite{hu} for the $n=1$ case (see \cite{nojiri8}).
The conditions (\ref{eq15b}) and (\ref{eq15d}) take the form
\be\label{eq17}
\lim_{R\rightarrow\infty}f(R)=-M^2\frac{\alpha}{\gamma}=-2\Lambda_i
\ee
\be\label{eq19}
\lim_{R\rightarrow 0}f(R)=-M^2\frac{\beta}{\delta}
\ee
But the restriction (\ref{eq15c}) can be solved exactly as
\be\label{eq19a}
R_0=\frac{\frac{2\Lambda}{M^2}\delta-\beta}{\alpha-\frac{2\Lambda}{M^2}\gamma}\sim\frac{\delta-\beta}{\alpha-\gamma}
\ee
where in the last approximation  we have assumed $M^2\approx2\Lambda$. This also imply that $\delta>\beta$ and $\alpha>\gamma$, and if we make $\delta=\beta$ then this leads to the approximation $R_0\sim0$ and $f(R_0)\sim f(0)\sim -M^2$.\\
To qualitatively evaluate the relationships between the parameters according to (\ref{eq17}) and (\ref{eq19}) we can write (\ref{eq16}) in more convenient form
\be\label{eq16a}
f(R)=-M^2\frac{\tilde{\alpha} (R/M^2)+\beta}{\tilde{\gamma} (R/M^2)+\delta},
\ee
where $\tilde{\alpha}=M^2\alpha$, $\tilde{\gamma}=M^2\gamma$ are dimensionless as well as  $\beta$ and $\delta$. Thus for instance if we assume $M\sim 10^{-33} {\rm ev}$, then in order to get  $\Lambda_i\sim 10^{46} {\rm ev}^2$  at large curvature in the early universe, the fraction 
$\tilde{\alpha}/\tilde{\gamma}$ should be of the order of $10^{112}$, while to meet the observations at late times, the fraction $\beta/\delta$ should be of the order of $1$. This leads to the relation 
\be\label{eq16b}
\frac{\tilde{\alpha}}{\tilde{\gamma}}>>\frac{\beta}{\delta}
\ee
which is equivalent to $\tilde{\alpha}\delta>>\beta\tilde{\gamma}$.
Thus the $f(R)$ correction in this model acts as an effective cosmological constant which is large at the beginning of the universe and becomes smaller at current epoch in the low-curvature universe.  The model also can pass the local gravity tests (earth or solar system) even using the restrictions (\ref{eq17}) and (\ref{eq19}). Setting $\tilde{\alpha}=1$ in  (\ref{eq16a}) we can rewrite the restrictions (\ref{eq17}) and (\ref{eq19}) respectively in the form
\be\label{eq16c}
\frac{1}{\tilde{\gamma}}=\frac{2\Lambda_i}{M^2},\,\,\,\,\,\, \delta=\beta
\ee
where we assumed as before $M\sim H_0\sim 10^{-33}  {\rm ev}$. Thus $f(R)$ can be written  as
\be\label{eq16d}
f(R)=-M^2\frac{\frac{R}{M^2}+\beta}{\frac{R}{\Lambda_i}+\beta}
\ee
and the relation (\ref{eq16b}) reduces to
\be\label{eq16e}
\tilde{\gamma}=\frac{M^2}{\Lambda_i}<<1
\ee
Applied to local gravity systems like the earth or the solar system, is clear that in this case $R$ is greater than $M^2$.  Taking (\ref{eq16e}) into account it follows that for curvatures below the $\Lambda_i$ value and for $\beta> R/M^2$ , we can neglect the first term in the denominator in (\ref{eq16d}), i.e. $R/\Lambda_i<<\beta$ (since $R/\Lambda_i<<R/M^2$). In these cases following approximation is valid
\be\label{eq16f}
f(R)\sim -M^2\frac{\frac{R}{M^2}+\beta}{\beta}\sim -M^2
\ee
For the earth $R\sim 10^{-50}{\rm ev}^2$ and $R/M^2\sim 10^{16}$. This indicates that in order to be consistent with the gravitational experiments on the earth, $\beta>10^{16}$ which validates the approximation  $f(R)\sim -M^2$. For the solar system $R\sim 10^{-61} {\rm ev}^2$ and therefore $R/M^2\sim 10^{5}$ which is much less than $\beta$ and also leads to (\ref{eq16f}). \\
Taking the derivatives in Eq. (\ref{eq16a}) one obtains
\be\label{eq24}
F'(R)=1-\frac{\tilde{\alpha}\delta-\beta\tilde{\gamma}}{(\tilde{\gamma} R/M^2+\delta)^2},\,\,\,\, F''(R)=2\frac{\tilde{\alpha}\delta-\beta\tilde{\gamma}}{(\tilde{\gamma} R/M^2+\delta)^3}\left(\frac{\tilde{\gamma}}{M^2}\right)
\ee
From this equation follows that the stability under small perturbations $F''(R)>0$ imply that $\tilde{\alpha}\delta-\beta\tilde{\gamma}=\beta(1-\tilde{\gamma})>0$ which is guaranteed by (\ref{eq16b}). And the condition for absence of ghosts (quantum stability) $F'(R)>0$ together with the condition of monotone decreasing on $F'(R)$, leads to 
\be\label{eq25}
(\tilde{\gamma} R/M^2+\delta)^2>(\tilde{\alpha}\delta-\beta\tilde{\gamma}),
\ee
the l.h.s. of this inequality increases with the increment of $R$, and its minimum value corresponding to $R=0$ is $\delta^2$. Therefore to satisfy this inequality is sufficient with
\be\label{eq26}
\delta^2>(\tilde{\alpha}\delta-\beta\tilde{\gamma}),\,\,\, {\rm or}\,\,\, \delta>\tilde{\alpha}
\ee
and using  (\ref{eq16c}) this last inequality is equivalent to the more simple restriction $\beta>1$, which is satisfied according to the constraint imposed on $\beta$ by local gravity (i.e. $\beta>10^{16}$). So the conditions for avoidance of ghosts and tachyon instabilities are widely satisfied.\\
Note that the parameter $\beta$ is important to ensure that the $f(R)$ correction satisfies the constraints imposed by local gravity systems. The parameter $\Lambda_i$ (or $\gamma$) is defined by the very early time behavior of the universe,  and it would be possible to restrict this parameter with good accuracy in the near future. Using the Eq. (\ref{eq24}) with the above restrictions on the parameters, is easy to check (whenever  $R<\Lambda_i$) that $f'(R)\sim -1/\beta$ and $F''(R)=f''(R)\sim 1/(\Lambda_i\beta^2)$.
\subsection*{Constant curvature solution.}
The model also contains constant curvature solutions in absence of matter including flat space-time. This last solution is obtained by setting $\beta\rightarrow 0$, leading to the result $f(0)=0$. 
Let's  consider the trace equation (\ref{eq3}) in absence of matter and look for solutions with constant Ricci curvature, i.e. the equation
\be\label{eq20}
R_cf'(R_c)-2f(R_c)-R_c=0
\ee
with f(R) given by (\ref{eq16}). Ignoring for simplicity the coefficient $-M^2$, the equation (\ref{eq16}) leads to a cubic algebraic equation in $R_c$ 
\be\label{eq21}
\frac{\gamma^2R_c^3+2\gamma(\alpha+\delta)R_c^2+\left(3\beta\gamma+\alpha\delta+\delta^2\right)R_c+2\beta\delta}{(\gamma R_c+\delta)^2}=0,
\ee
tat has one real root given by
\be\label{eq22}
R_c=\frac{1}{3\gamma}\left[-2(\alpha+\delta)+\frac{9\beta\gamma-5\alpha\delta-4\alpha^2-\delta^2}{A^{1/3}}-A^{1/3}\right]
\ee
where the dimensionless quantity $A$ is given by
\be\label{eq23}
\begin{aligned}
A=& 8\alpha^3-27\alpha\beta\gamma+15\alpha^2\delta+6\alpha\delta^2-\delta^3\\
&+\sqrt{\left(27\alpha\beta\gamma-8\alpha^3-15\alpha^2\delta-6\alpha\delta^2+\delta^3\right)^2-\left(4\alpha^2-9\beta\gamma+5\alpha\delta+\delta^2\right)^3}
\end{aligned}
\ee
The solution $R_c$ can take positive or negative values, leading to exact de Sitter or anti-de Sitter vacuum solutions and also to Schwarzschild-de Sitter(anti-de Sitter) depending on the geometry of the space-time.
Numerical calculations show that this solution takes positive values for certain relations between the signs of the parameters (taking into account the symmetry under $(\alpha,\beta,\gamma,\delta)\rightarrow (-\alpha,-\beta,-\gamma,-\delta))$, namely the necessary (but not sufficient) condition for positive $R_c$ is $sign(\alpha)=sign(\delta)=-sign(\beta)=-sign(\gamma)$ or $sign(\alpha)=sign(\beta)=-sign(\gamma)=-sign(\delta)$, or if one of the parameters has different sign than the others. 
Taking into account the condition of stability $m_F^2>0$ (see Eq. (\ref{eq9}) for the solution $R_c$, it can be satisfied in the case $sign(\alpha)=sign(\beta)=-sign(\gamma)=-sign(\delta)$ (this correlation of signs is in agreement with the expression (\ref{eq16})), which also guarantees that $f(R)$ does not have singularities due to zeros in the denominator. From (\ref{eq22}) follows that the flat space time solution can be obtained if $\beta=0$.
\subsection*{The scalar-tensor correspondence.}
After the conformal transformation of the metric by using the function $F'(R)$ (called the scalaron field), the initial action (\ref{eq1}) takes the form of the Einstein's term plus a scalar field non-minimally coupled to matter, as given in (\ref{eq14}). In what follows we will not consider the mater term (${\cal L}_m=0$). For the model (\ref{eq16a}) and using (\ref{eq13} ) we can find the curvature in terms of the scalar field from the relation (taking into account that $0<F'(R)<1$)
\be\label{eq27}
F'(R)=1+f'(R)=e^{-\sqrt{\frac{2}{3}}\kappa\phi},
\ee
giving
\be\label{eq28}
R=\frac{1}{\gamma}\left[M\left(\frac{\alpha\delta-\beta\gamma}{1-e^{-\sqrt{\frac{2}{3}}\kappa\phi}}\right)^{1/2}-\delta\right]
\ee
which leads to the "scalaron" potential
\be\label{eq29}
V(\phi)=\frac{1}{2\kappa^2\gamma}\left[\alpha M^2+\delta-\delta e^{-\sqrt{\frac{2}{3}}\kappa\phi}+2M\sqrt{(\alpha\delta-\beta\gamma)(1-e^{-\sqrt{\frac{2}{3}}\kappa\phi}})\right]e^{2\sqrt{\frac{2}{3}}\kappa\phi}
\ee
using (\ref{eq13}) and taking the first and second derivatives of the potential we find 
\be\label{eq30}
\frac{\partial V}{\partial \phi}=-\frac{1}{\sqrt{6}\kappa}\frac{2F(R)-RF'(R)}{F'(R)^2},
\ee
\be\label{eq31}
\frac{\partial^2V}{\partial\phi^2}=\frac{1}{3F''(R)}\left(1+\frac{RF''(R)}{F'(R)}-\frac{4F(R)F''(R)}{F'(R)^2}\right)
\ee
Note that the minimum of the potential is the same as discussed in the Jordan frame. Taking into account the minimum of potential (\ref{eq28}) into the second derivative we find the mass of the scalaron in the Einstein frame
\be\label{eq32}
m_{\phi}^2=\frac{1}{3F''(R)}\left(1-\frac{RF''(R)}{F'(R)}\right)=\frac{1}{3F'(R)}\left(\frac{F'(R)}{F''(R)}-R\right)
\ee
note that compared to the Einstein frame the mass in the Jordan frame is $m_F=\sqrt{F'}m_{\phi}$. Applied to the present  $F(R)$ model one finds for $m_{\phi}$
\be\label{eq33}
m_{\phi}^2=\frac{\left[(\gamma R+\delta)^3-M^2(\alpha\delta-\beta\gamma)(3\gamma R+\delta)\right](\gamma R+\delta)^2}{6\gamma M^2(\alpha\delta-\beta\gamma)\left[(\gamma R+\delta)^2-M^2(\alpha\delta-\beta\gamma)\right]}
\ee
we can qualitatively analyze the conditions for $m_{\phi}^2>0$ as follows: the expression in the square bracket in the denominator is positive if $(\gamma R+\delta)^2>M^2(\alpha\delta-\beta\gamma)$ which is the same condition (\ref{eq25}) that is satisfied by the inequality (\ref{eq26}). The numerator is positive if $(\gamma R+\delta)^3>M^2(\alpha\delta-\beta\gamma)(3\gamma R+\delta)$, which is satisfied if $(\gamma R+\delta)^2>3M^2(\alpha\delta-\beta\gamma)$, and to satisfy this inequality is sufficient with $\delta>3M^2\alpha$ (here we used (\ref{eq16b})). So the condition for absence of ghosts ($F'(R)>0$)  and of monotone decreasing of $F'(R)$  (\ref{eq26}) together with the condition $m_{\phi}^2>0$ is satisfied by $\delta>3M^2\alpha$. Note that this condition also holds for $m_F^2$ in the Jordan frame. Using (\ref{eq16c}) and the restriction on $\beta$ imposed by local gravity tests, we find that the inequality  $\delta>3M^2\alpha$
is equivalent to $\beta>3$ which is satisfied by the value of $\beta$ we found before. Analyzing the mass of the scalaron given by (\ref{eq31}) and under the conditions discussed before for the local systems, we find from (\ref{eq31}) that 
\be\label{eq34}
m_{\phi}^2\sim \Lambda_i\beta^2\left(1-\frac{R}{\Lambda_i\beta}\right)\sim \Lambda_i\beta^2
\ee
In order to avoid measurable corrections to the Newton law $m_{\phi}$ should be very large (the correlation length should be much smaller than the typical size of the system). Under the assumption that $\Lambda_i\sim 10^{46} {\rm ev}^2$ and as quoted before $\beta>10^{16}$, we find that $m_{\phi}$ is large enough to avoid corrections to the Newton law.
\subsection*{Matter stability}
An important test for any modified gravity model (specially if the model has not flat Minkowski space-time solution) is the matter stability under small perturbations of the scalar curvature created by matter in the Einstein approximation \cite{....}. In other words, we assume as the background curvature ($R_b$) the curvature created by a local system in the general relativity approximation and study the behavior of the linear perturbations  of this curvature $\delta R$. This means that the r.h.s. in Eq. (\ref{eq3}) becomes $\kappa^2T^{m}=-R_b$ and the Eq. (\ref{eq3}) takes the form
\be\label{eq34a}
RF'(R)-2F(R)+3F'''(R)\nabla_{\mu}R\nabla^{\mu}R+3F''(R)\Box R+R_b=0
\ee
Writing the scalar curvature in the form $R=R_b+\delta R$  ($\delta R<<R_b$) and neglecting the spatial dependence in Eq. (\ref{eq34a}) we come to the equation for $\delta R$ as
\be\label{eq35}
\frac{\partial^2 \delta R}{\partial t^2} +P(R_b)\delta R+Q(R_b)=0
\ee
where $P(R_b)$ and $Q(R_b)$ are "constants" that depend on the background curvature with $P(R_b)$  given by
\be\label{eq36}
\begin{aligned}
P(R_b)=&\left(\frac{F'''(R_b)^2}{F''(R_b)^2}-\frac{F^{IV}(R_b)}{F''(R_b)}\right)\nabla_{\mu}R_b\nabla^{\mu}R_b+\frac{F'(R_b)}{3F''(R_b)}\\&-\frac{R_b}{3}+\frac{R_bF'(R_b)F'''(R_b)}{3F''(R_b)^2}-\frac{2F(R_b)F'''(R_b)}{3F''(R_b)^2}+\frac{R_bF'''(R_b)}{3F''(R_b)^2}
\end{aligned}
\ee
then the condition for stability of the system is resumed in $P(R_b)>0$. Applied to the model (\ref{eq16d}) and using the restrictions we have discussed for local systems, we find
\be\label{eq37}
\begin{aligned}
&F'(R_b)= 1-\frac{1}{\beta}\sim 1,\;\;\;\;\; F''(R_b)= \frac{2\gamma\beta(1-\tilde{\gamma})}{(\gamma R_b+\beta)^3}\sim \frac{1}{\Lambda_i\beta^2},\\
&F'''(R_b)= -\frac{6\gamma^2\beta(1-\tilde{\gamma})}{(\gamma R_b+\beta)^4}\sim -\frac{1}{\Lambda_i^2\beta^3},\;\; F^{IV}(R_b)= 
\frac{24\gamma^3\beta(1-\tilde{\gamma})}{(\gamma R_b+\beta)^5}\sim \frac{1}{\Lambda_i^3\beta^4}
\end{aligned}
\ee
which leads to
\be\label{eq38}
P(R_b)\approx \frac{1}{3}\Lambda_i\beta^2-\frac{1}{3}R_b\sim \frac{1}{3}\Lambda_i\beta^2>0
\ee
where we used the natural assumption $\Lambda_i>>R_b$. Therefore the curvature produced by local gravity systems in the Einstein approximation is stable under small perturbations in the modified gravity model (\ref{eq16d}).
\section{Discussion}
We proposed a model of modified gravity $f(R)$ that describes both the early time inflation and late-time acceleration. For large curvature $f(R)$ becomes almost constant and plays the role of effective cosmological constant that generates inflation and for small (in the case $\beta\ne 0$) curvature $f(R)$ becomes again almost constant being interpreted as the cosmological constant that leads to the current accelerated expansion of the universe. The proposed model contains some of the models already studied in the literature as limit cases. Thus the case $\gamma=0, \beta=0$ gives the power-law model while $\alpha=\delta=0$ gives the inverse power-law model, $\beta=0, \eta=1$ coincides with the Hu-Sawicki model for $n=1$ \cite{hu} and the limit $\alpha=\gamma=0$ leads to the $\Lambda$CDM model.  Concerning the viability, the model presents several attractive facts as it satisfies the conditions of absence of ghost graviton (antigravity) $F'(R)>0$, the absence of tachyon instabilities $F''(R)>0$, which lead to stable de Sitter solutions. For local systems it is possible to find large enough scalaron mass in order to avoid detectable corrections to the Newton's law. The model is also stable under small perturbations of the constant curvature produced by local gravity systems (matter stability).\\
Resuming, we presented a model that satisfies the standard conditions of viability and describes the current observed accelerated expansion without entering in conflict with local tests of gravity. In this model the inflationary period is explained by an effective cosmological constant at the early stage at large curvature. As the universe evolves the curvature continues decreasing and the $f(R)$ correction tends to small effective cosmological constant
that leads to the current accelerated expansion. For the sample of values and correlations between the parameters that we considered here, the correction to the Einstein-Hilbert gravitational action becomes relevant at both, early and late universe.
\section*{Acknowledgments}
This work was supported by Universidad del Valle under projects  CI 7883 and CI 7890.

\end{document}